\input amstex
\magnification=1200
\documentstyle{amsppt}
\NoRunningHeads
\NoBlackBoxes
\topmatter
\title Quantum string field theory and psychophysics\endtitle
\author Denis V. Juriev\endauthor
\affil ul.Miklukho-Maklaya 20-180, Moscow 117437 Russia\linebreak
(e-mail: denis\@juriev.msk.ru)\endaffil
\date physics/0008058\enddate
\abstract\nofrills The quantum string field theoretic structure of interactive 
phenomena is discussed.
\endabstract
\endtopmatter
\document
This note continues the author's researches on the boundary of experimental
mathematics, psychophysics and computer science, which were initiated about
ten years ago. Precisely, it is devoted to the unraveling of quantum string
field theoretic (general aspects of this theory are discussed in the book [1]
and its mathematical formalism based on the infinite dimensional geometry 
is exposed in [2]) structures in the picture described in two previous 
notes [3]. The results may be significant for the constructing of a very
important bridge between fundamental theoretical high-energy physics and 
modern psychophysics. The interactive game theoretic surrounding of the 
least may essentially enrich the quantum string field theory by new original
features, which will be interesting for pure mathematicians. Such alliance
may be interesting to the theoretical physicists as supplying their 
sophisticated constructions with a very simple and inexpensive experimental 
verification.

\head I. Interactive phenomena: experimental detection and analysis [3]\endhead

\subhead 1.1. Experimental detection of interactive phenomena\endsubhead
Let us consider a natural, behavioral, social or economical system $\Cal S$.
It will be described by a set $\{\varphi\}$ of quntities, which characterize
it at any moment of time $t$ (so that $\varphi=\varphi_t$). One may suppose
that the evolution of the system is described by a differential equation 
$$\dot\varphi=\Phi(\varphi)$$
and look for the explicit form of the function $\Phi$ from the experimental
data on the system $\Cal S$. However, the function $\Phi$ may depend on time,
it means that there are some hidden parameters, which control the system
$\Cal S$ and its evolution is of the form
$$\dot\varphi=\Phi(\varphi,u),$$
where $u$ are such parameters of unknown nature. One may suspect that such 
parameters are chosen in a way to minimize some goal function $K$, which may 
be an integrodifferential functional of $\varphi_t$:
$$K=K(\left[\varphi_{\tau}\right]_{\tau\le t})$$
(such integrodifferential dependence will be briefly notated as 
$K=K([\varphi])$ below). More generally, the parameters $u$ may be divided
on parts $u=(u_1,\ldots,u_n)$ and each part $u_i$ has its own goal function
$K_i$. However, this hypothesis may be confirmed by the experiment very 
rarely. In the most cases the choice of parameters $u$ will seem accidental
or even random. Nevertheless, one may suspect that the controls $u_i$ are 
{\sl interactive}, it means that they are the couplings of the pure controls 
$u_i^\circ$ with the {\sl unknown or incompletely known\/} feedbacks:
$$u_i=u_i(u_i^\circ,[\varphi])$$
and each pure control has its own goal function $K_i$. Thus, it is
suspected that the system $\Cal S$ realizes an {\sl interactive game}.
There are several ways to define the pure controls $u_i^\circ$. One of them
is the integrodifferential filtration of the controls $u_i$:
$$u^\circ_i=F_i([u_i],[\varphi]).$$
To verify the formulated hypothesis and to find the explicit form of the
convenient filtrations $F_i$ and goal functions $K_i$ one should use the
theory of interactive games, which supplies us by the predictions of the
game, and compare the predictions with the real history of the game for
any considered $F_i$ and $K_i$ and choose such filtrations and goal functions,
which describe the reality better. One may suspect that the dependence of
$u_i$ on $\varphi$ is purely differential for simplicity or to introduce the
so-called {\sl intention fields}, which allow to consider any interactive
game as differential. Moreover, one may suppose that
$$u_i=u_i(u_i^\circ,\varphi)$$
and apply the elaborated procedures of {\sl a posteriori\/} analysis and
predictions to the system.

In many cases this simple algorithm effectively unravels the hidden 
interactivity of a complex system. However, more sophisticated procedures
exist [3].

Below we shall consider the complex systems $\Cal S$, which have been yet
represented as the $n$-person interactive games by the procedure described
above. 

\subhead 1.2. Functional analysis of interactive phenomena\endsubhead
To perform an analysis of the interactive control let us note that often for 
the $n$-person interactive game the interactive controls 
$u_i=u_i(u_i^\circ,[\varphi])$ may be represented in the form 
$$u_i=u_i(u_i^\circ,[\varphi];\varepsilon_i),$$
where the dependence of the interactive controls on the arguments
$u_i^\circ$, $[\varphi]$ and $\varepsilon_i$ is known but the 
$\varepsilon$-parameters $\varepsilon_i$ are the unknown or incompletely
known functions of $u_i^\circ$, $[\varepsilon]$. Such representation is
very useful in the theory of interactive games and is called the 
{\sl $\varepsilon$-representation}. 

One may regard $\varepsilon$-parameters as new magnitudes, which characterize
the system, and apply the algorithm of the unraveling of interactivity to
them. Note that $\varepsilon$-parameters are of an existential nature 
depending as on the states $\varphi$ of the system $\Cal S$ as on the
controls. 

The $\varepsilon$-parameters are useful for the functional analysis of
the interactive controls described below.

First of all, let us consider new integrodifferential filtrations $V_\alpha$:
$$v^\circ_\alpha=V_\alpha([\varepsilon],[\varphi]),$$
where $\varepsilon=(\varepsilon_1,\ldots,\varepsilon_n)$. 
Second, we shall suppose that the $\varepsilon$-parameters are expressed via 
the new controls $v^\circ_\alpha$, which will be called {\it desires:}
$$\varepsilon_i=\varepsilon(v^\circ_1,\ldots,v^\circ_m,[\varphi])$$
and the least have the goal functions $L_\alpha$. The procedure of unraveling
of interactivity specifies as the filtrations $V_\alpha$ as the goal functions 
$L_\alpha$.

\subhead 1.3. SD-transform and SD-pairs\endsubhead
The interesting feature of the proposed description (which will be called the
{\it S-picture}\/) of an interactive system $\Cal S$ is that it contains as 
the real (usually personal) subjects with the pure controls $u_i$ as the 
impersonal desires $v_\alpha$. The least are interpreted as certain 
perturbations of the first so the subjects act in the system by the 
interactive controls $u_i$ whereas the desires are hidden in their actions. 

One is able to construct the dual picture (the {\sl D-picture\/}),
where the desires act in the system $\Cal S$ interactively and the
pure controls of the real subjects are hidden in their actions.
Precisely, the evolution of the system is governed by the equations
$$\dot\varphi=\tilde\Phi(\varphi,v),$$
where $v=(v_1,\ldots,v_m)$ are the $\varepsilon$-represented interactive 
desires:
$$v_\alpha=v_\alpha(v^\circ_\alpha,[\varphi];\tilde\varepsilon_\alpha)$$
and the $\varepsilon$-parameters $\tilde\varepsilon$ are the unknown or
incompletely known functions of the states $[\varphi]$ and the pure
controls $u_i^\circ$.

D-picture is convenient for a description of systems $\Cal S$ with a
variable number of acting persons. Addition of a new person does not
make any influence on the evolution equations, a subsidiary term to
the $\varepsilon$-parameters should be added only.

The transition from the S-picture to the D-picture is called the
{\it SD-transform}. The {\it SD-pair\/} is defined by the evolution
equations in the system $\Cal S$ of the form
$$\dot\varphi=\Phi(\varphi,u)=\tilde\Phi(\varphi,v),$$
where $u=(u_1,\ldots,u_n)$, $v=(v_1,\ldots,v_m)$, 
$$\aligned
u_i=&u_i(u_i^\circ,[\varphi];\varepsilon_i)\\
v_\alpha=&v_\alpha(v^\circ_\alpha,[\varphi];\tilde\varepsilon_\alpha)
\endaligned$$
and the $\varepsilon$-parameters $\varepsilon=(\varepsilon_1,\ldots,
\varepsilon_n)$ and $\tilde\varepsilon=(\tilde\varepsilon_1,\ldots,
\tilde\varepsilon_m)$ are the unknown or incompletely known functions of
$[\varphi]$ and $v^\circ=(v^\circ_1,\ldots,v^\circ_m)$ or
$u^\circ=(u^\circ_1,\ldots,u^\circ_n)$, respectively. 

Note that the S-picture and the D-picture may be regarded as complementary
in the N.Bohr sense. Both descriptions of the system $\Cal S$ can not be 
applied to it simultaneously during its analysis, however, they are compatible 
and the structure of SD-pair is a manifestation of their compatibility.

\head II. Quantum string field theoretic structure of interactive 
phenomena\endhead

\subhead 2.1. The second quantization of desires\endsubhead
Intuitively it is reasonable to consider systems with a variable number
of desires. It can be done via the second quantization. 

To perform the second quantization of desires let us mention that they
are defined as the integrodifferential functionals of $\varphi$ and
$\varepsilon$ via the integrodifferential filtrations. So one is able
to define the linear space $H$ of all filtrations (regarded as classical 
fields) and a submanifold $M$ of the dual $H^*$ so that $H$ is naturally
identified with a subspace of the linear space $\Cal O(M)$ of smooth functions
on $M$. The quantized fields of desires are certain operators in the
space $\Cal O(M)$ (one is able to regard them as unbounded operators in its
certain Hilbert completion). The creation/annihilation operators are
constructed from the operators of multiplication on an element of $H\subset
\Cal O(M)$ and their conjugates.

To define the quantum dynamics one should separate the quick and slow time.
Quick time is used to make a filtration and the dynamics is realized in
slow time. Such dynamics may have a Hamiltonian form being governed by
a quantum Hamiltonian, which is usually differential operator in $\Cal O(M)$.

If $M$ coincides with the whole $H^*$ then the quadratic part of a Hamiltonian
describes a propagator of the quantum desire whereas the highest terms
correspond to the vertex structure of self-interaction of the quantum field. 
If the submanifold $M$ is nonlinear the extraction of propagators and 
interaction vertices is not straightforward.

\subhead 2.2. Quantum string field theoretic structure of the second 
quantization of desires\endsubhead
First of all, let us mark that the functions $\varphi(\tau)$ and
$\varepsilon(\tau)$ may be regarded formally as an open string. The target 
space is a product of the spaces of states and $\varepsilon$-parameters. 

Second, let us consider a classical counterpart of the evolution of the 
integrodifferential filtration. It is natural to suspect that such evolution
is local in time, i.e. filtrations do not enlarge their support (as a time
interval) during their evolution. For instance, if the integradifferential
filtration depends on the values of $\varphi(\tau)$, $\varepsilon(\tau)$
for $\tau\in[t_0-t_1,t_0-t_2]$ at the fixed moment $t_0$ it will depend
on the same values for $\tau\in[t-t_1,t-t_2]$ at other moments $t>t_0$.
This supposition provides the reparametrization invariance of the classical
evolution. Hence, it is reasonable to think that the quantum evolution is
also reparametrization invariant. 

Reparametrization invariance allows to apply the quantum string field 
theoretic models to the second quantization of desires. For instance, one
may use the string field actions constructed from the closed string vertices
(note that the phase space for an open string coincides with the configuration
space of a closed string) or string field theoretic nonperturbative actions.
In the least case the theoretic presence of additional "vacua" (minimums
of the string field action) is very interesting.

\subhead 2.3. Additional fields and virtual subjects\endsubhead
Often quantum string field theory claims an introduction of additional
fields (such as bosonised ghosts). Let us consider such fields in the 
D-picture. 

In D-picture desires have their own $\varepsilon$-parameters and depend
on the pure controls of subjects. These pure controls may be obtained
from the $\varepsilon$-parameters of desires via integrodifferential
filtrations. One is able to apply such filtrations to the additional
fields. There are two possibilities. First, the result is expressed
via the known pure controls. Second, the result is a new pure control of
a {\it virtual subject}. Certainly, any experimental detection of virtual
subjects is extremely interesting.

\head III. Conclusions \endhead

Thus, the quantum string field theoretic structure of interactive phenomena
is described. Possible qualitative effects, which are produced by this
structure and confirm its presence, are emphasized. Perspectives are
briefly specified.

\Refs
\roster
\item"[1]" Green M.B., Schwarz J.H., Witten E., Superstring theory. Cambridge
Univ.Press, Cambridge, 1988.
\item"[2]" Juriev D., Infinite dimensional geometry and quantum field theory
of strings. I-III. AGG 11(1994) 145-179 [hep-th/9403068], RJMP 4(3) (1996)
187-314 [hep-th/9403148], JGP 16 (1995) 275-300 [hep-th/9401026]; String
field theory and quantum groups. I: q-alg/9708009.
\item"[3]" Juriev D., Experimental detection of interactive phenomena and
their analysis:\linebreak math.GM/0003001; New mathematical methods for 
psychophysical filtering of experimental data and their processing: 
math.GM/0005275.
\endroster
\endRefs
\enddocument